# Dynamic Radar Networks of UAVs

Anna Guerra, University of Bologna, Davide Dardari, University of Bologna, Petar M. Djurić, Stony Brook University

*In the incoming years, the low-aerial space will be crowded by unmanned aerial vehicles (UAVs), which will be providing different services. In this expected context, an emerging problem is to detect and track unauthorized or malicious mini/micro UAVs. In contrast to current solutions mainly based on fixed terrestrial radars, this article will put forth the idea of a dynamic radar network (DRN) composed of UAVs able to smartly adapt their formation-navigation control to best track malicious UAVs in real-time, with high accuracy, and in a distributed fashion. To this end, some technological solutions and the main methods for target detection and tracking will be described. Further, an optimized navigation scheme will be developed according to an information-seeking approach. Some examples of simulation results and future directions of work will be finally presented highlighting the advantages of dynamic and reconfigurable networks over static ones.*

## Motivations

The use of civil unmanned aerial vehicles (UAVs) in densely inhabited areas like cities is expected to open an unimaginable set of new applications thanks to their low-cost and high flexibility. They can be used for enabling smart services in low-altitude air space (below 150 m), as for instance goods delivery (e.g., Amazon prime air), taxi drones (e.g., Uber air), and monitoring, according to the U-Space roadmap [1], [2].

As an example, UAVs have been recently proposed as a complementary aerial platform in 5G cellular networks to enhance communication services for terrestrial users thanks to their capability of reacting to the fast variations of traffic demand, and to rely on dominant line-of-sight (LOS) links [3].

At the same time, the idea of having swarms of UAVs in future cities might be accepted with difficulty by the public because of their potential malicious use. Indeed, UAVs, for example, can hide behind buildings for criminal activities like terrorist attacks, or can inhibit the functionality of authorized UAV networks [4].

Currently, UAV safety and security solutions rely on communication with the air traffic management (ATM) infrastructure and/or ad hoc fixed on-ground radars. On the one hand, the ATM processes the information acquired and provided by UAVs (e.g., the e-identification packet containing the UAV ID and position, as foreseen by the U1 services of the U-space [2]), which can be easily subjected to cybersecurity attacks.

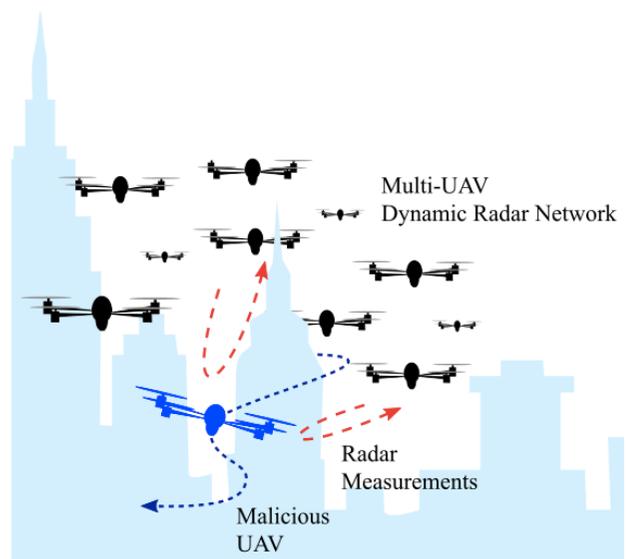

*Figure 1.* Pictorial representation of a DRN considered in this paper.

On the other hand, to improve the safety and security level, one can deploy ad-hoc terrestrial radar systems. Typically, terrestrial radars for UAV detection are of two types: fixed radar systems with operating ranges of about 500-2000 meters in open space [5], and fixed systems based on RF, vision or acoustic sensors, mostly of small sizes and used in critical areas (e.g., airports) [6]. Another recent solution is to use low-power passive radars that exploit signals of opportunities (e.g., cellular, Wi-Fi signals) to illuminate the objects in the surroundings and process the signal backscattered by UAVs [7].

Unfortunately, all these systems could fail in harsh environments like cities due to obstacles that prevent the reception of the signal by terrestrial radars, and, consequently, will increase missed detection and inaccuracy of tracking malicious UAVs. Moreover, the deployment of ad-hoc terrestrial

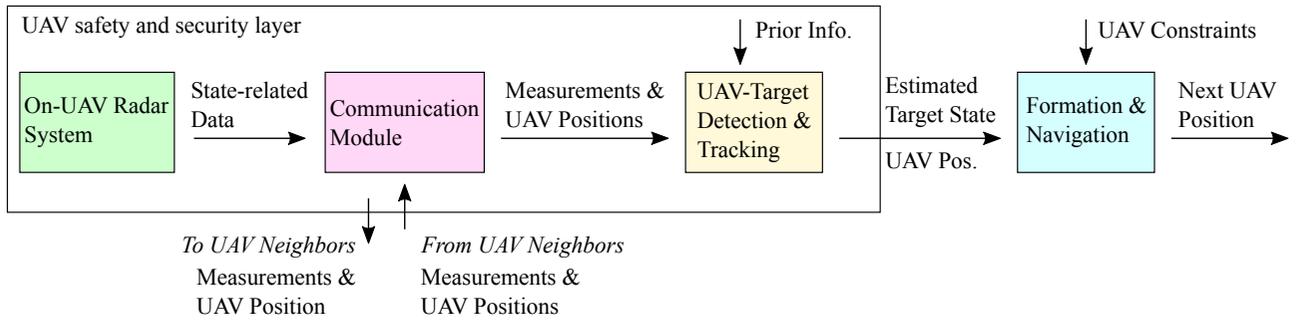

*Figure 2. A block diagram for decentralized joint detection and tracking performed at each UAV.*

radar or vision-based infrastructures might not be always feasible or economically sustainable.

For the above-mentioned reasons, the problem of fast, reliable, and autonomous detection and tracking of malicious UAVs is challenging and still an unsolved issue.

In this context, one approach to address this problem is to use "UAVs to monitor UAVs" (patrolling drones), and, in this article, we illustrate and elaborate on the possibility of adopting a UAV-based dynamic radar network (DRN). A DRN is a network of UAVs acting as a cooperative distributed radar sensing system for real-time high-accuracy tracking of non-authorized UAVs (malicious targets). Figure 1 provides a pictorial representation of a DRN.

This system will be able to observe the environment from different privileged viewpoints compared to on-ground systems. In fact, UAV-radars will fly at a certain altitude so that it will be possible to detect and track targets with an unprecedented level of accuracy. Moreover, thanks to UAV-to-UAV collaboration [8], the sensing capability and the capacity to optimize the trajectory in real time of each individual UAV will be augmented.

In the following, the problem of a joint autonomous navigation and target tracking problem will be analyzed considering the position of sensors being optimized through a formation control algorithm. In this sense, due to lack of prior knowledge of the environment and of the target trajectory, off-line path planning algorithms [9] will not be efficient, and, hence, a UAV-radar should interact with nearby UAV-radars in order to retrieve sufficient information for planning its optimized trajectories on-the-fly. In order to meet the low latency constraint for tracking and navigation as well as reliability against cyber-attacks, a fully distributed scheme, where data are exchanged via multi-hops, is considered, where the UAVs rely only on the information exchanged with their first-tier neighbors.

The final goal is to demonstrate the feasibility of the DNRs for enhancing the safety and security of low-altitude air space as well as to propose DNRs as a more reliable alternative (or complementary) to vision- or terrestrial-based solutions in the presence of bad weather conditions or several obstructions.

## Problem Statement

A DRN of UAVs can be described as a set of mobile reference nodes (i.e., with a-priori known positions, for instance available from GPS signals) that navigate in an outdoor environment. UAVs are equipped with radar sensors that provide measurements used for detecting and tracking, for instance, the position and velocity of a malicious passive (non-cooperative) UAV, in the following referred to as target. If a target is detected, a tracking process is started whose purpose is to minimize the tracking error through a suitable formation-navigation control algorithm and, consequently, to enable safety countermeasures.

For this purpose, each UAV acquires radar measurements and exchanges them with neighbors via multi-hops (*distributed network*), in order to lower the latency of communications. According to this distributed topology, each UAV can be considered as a central unit capable of locally assessing the situation, detect and track the non-authorized target, and taking navigation decisions accordingly, in an autonomous way. This improves also the reliability of the network vis-à-vis UAV failures or external attacks. Hence, in accordance with Fig. 2, each UAV performs the following steps:

1) The first task focuses on retrieving useful information from radar measurements, i.e., from the signal backscattered by the environment where the malicious target navigates. For example, Doppler shift, ranging and/or bearing information are extrapolated from the backscattered signal in order to be processed by the detection and tracking algorithm.

2) Once the UAV has acquired its own measurements, it communicates this information to the neighbors together with its own position, and it receives back the same data from neighboring UAVs via multi-hop propagation.

3) Given the measurements and the positions of the other UAVs, the presence of a malicious target can be detected and its state can be tracked. For example, a Bayesian estimator can be used to compute the a-posteriori probability distribution of the state given the history of measurements, as it will be detailed in the sequel.

4) The last step is the UAV position update that will allow the UAV to reach its next position according to a given task (control law). Since the quality of the measurements depends on the DRN geometry and target position, the control law should properly dynamically change the UAV formation and position in order to maximize the quality of the tracking process.

## Methods

The main phases for detection and tracking of a target by a DRN will be discussed in this section, with a particular emphasis on UAV-compliant solutions.

### UAV-Radar Measurements

Several types of sensors can be mounted on UAVs such as cameras, RF sensors, microphones, and radars. One advantage of radars compared to vision-based sensors is that their performance is not affected by lighting or weather conditions, and they are less sensitive to non-line-of-sight (NLOS) conditions.

Radar systems transmit a signal that objects and targets present in the environment may backscatter based on their capability to reflect electromagnetic waves. This latter property is measured by the radar cross section (RCS) of the targets, and can impact on the final performance. Starting from the backscattered signal, a radar can estimate range, but also angle-of-arrival, and velocity of the targets.

In particular, ranging (distance estimation) can be performed starting from the received signal strength (RSS), using suitable path-loss models, or it can be obtained by estimating the time-of-flight (TOF) of the signal reflected by the target. The main advantage of performing RSS estimates is the low-complexity and low-cost of the needed technology. However, these estimates are extremely sensitive to multipath, and their accuracy also deteriorates when the distance between nodes increases. On the contrary, TOF-based measurements can be more accurate, especially if wideband signals are adopted.

Another possibility is to estimate the angle-of-arrival (AOA) by adopting multiple antennas or by exploiting ad-hoc UAV rotations of directive antennas.

Finally, Doppler-shifts measurements can be useful for estimating the target velocity, and they can be inferred through frequency or phase estimation.

In all cases, the measurements are subjected to uncertainties, related to noise, multipath, interference, etc., that can be mitigated by a proper choice of the underlaying technologies. In this sense, a promising radar technology for UAVs is the frequency modulated continuous wave (FMCW) radar. Differently from pulse radars that transmit short pulses periodically, FMCW radars interrogate the environment with a signal linearly modulated in frequency (namely, chirp). Sometimes, in order to measure multiple targets, multiple chirps are transmitted in a fixed time window (chirp train). Once the signal is received back by the radar, it is combined with a template of the transmitted waveform by a mixer. As a result, different target-related parameters, such as ranging and Doppler shifts, can be inferred by processing the frequency and phase information of the signal at the output of this mixer.

More specifically, each chirp presents a delay proportional to the distance from the targets that can be estimated using a two-dimensional Fourier transform (range-FFT) able to separate between different beat frequencies (one for each target) in the spectrum. In the same manner, to retrieve velocity information, it is possible to rely on phase differences between different received chirps, or, directly, on Doppler-shift estimates. Finally, if the FMCW radar consists of multiple transmitting and receiving antennas (MIMO radar), also AOA can be estimated through the measurement of the phase difference between antennas.

A promising solution is to operate at millimeter-waves so that the FMCW radars can be miniaturized and equipped with multiple antennas. For example, working at 77 GHz will permit to have a resolution smaller than a millimeter thanks to the higher available bandwidth (up to 4 GHz). Moreover, by lowering the wavelength, the integration of multiple antennas in smaller space (small form factor) will be possible as well as a reduction of weight of the payload. Example of FMCW for UAVs can be found in [10] and the references therein.

### Detection Problem

The detection of a malicious UAV is a very challenging research problem, especially when dealing with mini/micro-UAVs, and, hence, in this section, we review the main detection techniques without entering into mathematical details. We refer the readers to [6] for more details on this subject.

Different methods can be envisioned for UAV detection, based on static systems, or on patrolling UAVs. It is possible to classify them based on the sensors adopted for the measurements, as follows:

- *RF-based detection:* In this case, low-cost and low-complexity RF sensors, as for example software defined radios (SDRs), can be mounted on UAVs and can be used to "sniff" the communications between the target UAV and its controller. The collected RF signals are usually processed by classifiers (e.g., using machine learning techniques) able to distinguish between RF signals emitted by UAVs. The disadvantage of this method is that a prior training phase is needed in order to identify different UAVs. Moreover, these techniques depend on the transmit power and receiver sensitivity, and they fail if the target is an autonomous UAV not remotely controlled, and, thus, not emitting any kind of signal.

- *Radar-based detection:* In this case, low-cost and lightweight FMCW radars can be used to detect targets in the surrounding environment. Differently from vision-based techniques, they perform well even in NLOS conditions and, if millimeter-waves are adopted, they also allow the estimation of the micro-Doppler signature from the energy backscattered by UAV propellers. Such signatures allow to classify drones, to detect the presence of additional payloads, and to distinguish them from other flying objects (e.g., birds). Because radar measurements are based on backscattered signals, the RCS of the targets can make the detection difficult, especially when dealing with micro drones [11].

- *Sound-based detection:* This method employs single or multiple microphones to detect the characteristic noise produced by UAVs when flying. A major drawback of the method is that it fails in urban or noisy areas, and it could not work efficiently if the microphones are installed on-board patrolling UAVs that emit a similar noise.
- *Vision-based detection:* This technique involves the use of camera sensors. They require LOS conditions to operate properly. Moreover, they encounter difficulties when the UAVs are flying in between buildings.

### *Tracking Problem*

Once the detection has been performed, the tracking goal is to estimate the state of the target (e.g., its position and velocity) starting from the collected measurements. In this article, we focus on statistical techniques that do not rely on geometric considerations or database/look-up tables (as for fingerprinting), but estimate the state of the target over time considering the history of measurements.

In this sense, Bayesian filtering methodologies, based on Kalman filtering (KF) or particle filtering (PF), have demonstrated to be powerful tools to solve the tracking problem thanks to their capabilities of dealing with heterogeneous measurements, statistical characterization of uncertainties, and target mobility models.

Within the Bayesian framework, the main goal is to estimate the full joint posterior probability of the state at time instant $k$, $s^{(k)}$, given measurements up to the current time instant, $z^{(1:k)}$. When the target state and the measurements form a Markov sequence, it is possible to define a probabilistic state-space Markovian model by considering the following three statistical models:

- *Prior information.* It represents the statistical description of the state at time instant 0, for instance at the output of the detection process;
- *Measurement model.* It describes how the state is related to the measurements by the likelihood distribution $p(z^{(k)} | s^{(k)})$;
- *Mobility model.* It describes how the state evolves in time by $p(s^{(k)} | s^{(k-1)})$.

Given this state-space model, the Bayesian filtering is a recursive approach that permits to estimate the marginal posterior distribution of the target state given the measurements. To this end, three steps are necessary [12]:

- *Initialization.* The marginal at time step 0 is set equal to the prior;
- *Prediction step.* By exploiting the mobility mode, it is possible to derive the predictive distribution of the state $p(s^{(k)} | z^{(1:k-1)})$;
- *Update step.* Once a new measurement becomes available, the marginal posterior of the state, i.e. $p(s^{(k)} | z^{(1:k)})$, can be computed by applying the Bayes' rule.

Finally, given the marginal posterior of the state, a point estimate $\hat{s}^{(k)}$ of $s^{(k)}$ can be derived, for example, adopting a minimum mean square error or a maximum a posteriori criterion.

In our investigated system, the observation functions are non-linear (e.g., Euclidean distance) and the observation noise is Gaussian. In this case, two practical methods are the Extended Kalman Filter (EKF) and the unscented KF (UKF), that provide a simple solution to the filtering equations.

### *UAV Path-Planning Problem*

In this section, we propose an autonomous control at each UAV designed to estimate its next waypoint (location) in order to maximize its capability to best track the target, considering the locations and measurements of the other UAVs.

The tracking performance mainly depends on the prior information acquired (if present), on the UAV network formation (geometry) and on the quality (uncertainties) of the collected measurements. Since the DRN is distributed, the optimization problem should be locally solved at each UAV based on information coming from multi-hop communications. For this reason, the final solution will be not optimal because UAVs can have only partial/different views of the full network geometry [13].

Path planning and optimization for UAVs has attracted much research attention over the years [9]. The optimization criterion is usually based on the minimization of the information cost that captures the quality of the tracking process. In this context, the cost function can be: (i) a metric assessing the performance of a specific tracking estimator (e.g., the state covariance matrix at the output of the EKF); (ii) a metric independent of the specific localization/tracking estimator, e.g., the Fisher Information Matrix (FIM) evaluated on the target location estimate.

To enforce "agnosticism" of the chosen estimator, next we describe the second approach. In this case, to implement the cost function, one can consider the optimal experimental design (OED): for example, the A-optimality minimizes the trace of the FIM inverse; the D-optimality considers the determinant, while the E-optimality minimizes the maximum eigenvalue of the inverse matrix [14].

With reference to the system and the problem previously described, each UAV solves an optimization problem to infer its control law based on the collected information. This problem can be formulated as,

$$\left(\ell_i^{(k+1)}\right)^* = \min_{\ell_i^{(k+1)}} \mathcal{C}\left(\mathrm{IM}_i^{(k)}\left(\ell_i^{(k+1)}, \hat{s}_i^{(k+1|k)}\right)\right), \quad (1)$$

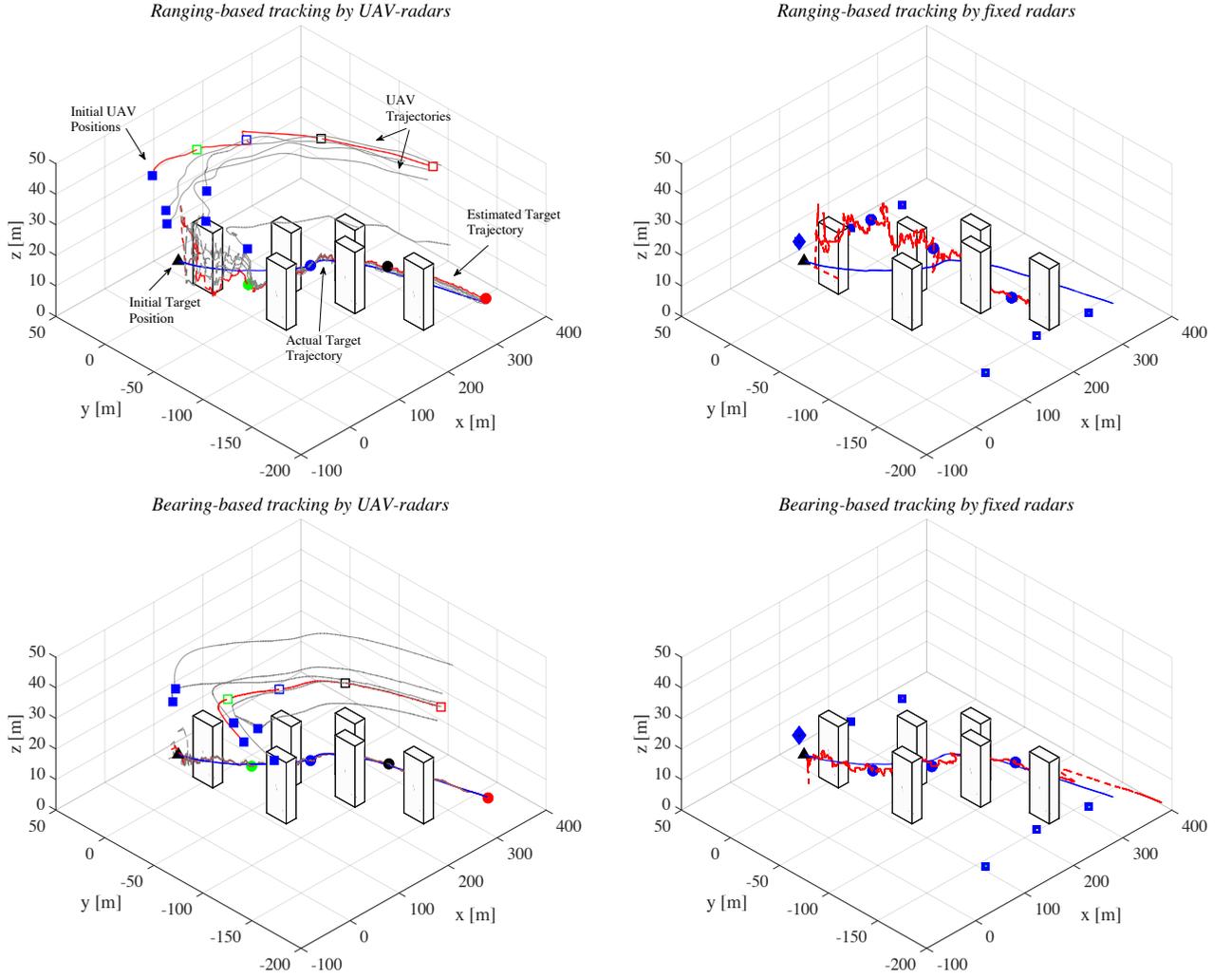

*Figure 3.* Simulation scenarios in presence of obstacles. On top, ranging-based tracking scenarios with a dynamic UAV radar network (left) and an ad-hoc fixed network (right). On the bottom, settings with bearing-based tracking measurements for dynamic and fixed radars.

where $\ell_i^{(k)}$ contains all the UAVs locations as known by the $i$-th UAV at time instant $k$, $\hat{s}_i^{(k+1|k)}$ is the predicted target state at the $i$-th UAV at time instant $k$, $\mathcal{C}(\cdot)$ is an optimal design cost function (e.g., the determinant) and $\text{IM}_i^{(k)}(\cdot)$ is an information metric capturing the accuracy of the tracking process, dependent on the UAV positions and on the target state (e.g., the inverse of the FIM). Because the information metric depends on the actual value of the state that is unknown, its expression is approximated by considering the estimate provided by the Kalman filter. Note that since the processing in DRNs is distributed, the measurements might be delayed by the number of hops between the UAVs and, hence, they may provide non-updated (i.e., aged) information about the target.

The UAV kinematic and anti-collision constraints, such as the maximum UAV turning rate or the distance to be kept from problem as non-linear inequality constraints.

Finally, according to the UAV transition model, the control signal at the $i$-th UAV can be found as $u_i^{(k)} = \left(\ell_i^{(k+1)}\right)^* - \ell_i^{(k)}$. When the UAVs arrive at their next positions, a new measurement and tracking phase is performed and the entire control process is repeated.

To solve the trajectory problem in (1), one can rely on an approach based on optimization theory (e.g., non-linear programming, dynamic programming, etc.) or on a more advanced approaches of machine learning (e.g., reinforcement learning algorithms).

Once the detection and the tracking of UAVs have been performed, several interdiction techniques can be undertaken to neutralize the threats posed by the malicious UAV. Some examples are reported in [6].

|  | RMSE on Target Position [m] | | RMSE on Target Velocity [m/s] | |
| --- | --- | --- | --- | --- |
|  | Fixed Radars | UAV Dynamic Radars | Fixed Radars | UAV Dynamic Radars |
| Ranging-only | 15.12 m | 3.94 m | 0.24 m/s | 0.18 m/s |
| Ranging & Doppler | 14.50 m | 2.58 m | 0.21 m/s | 0.14 m/s |
| Bearing-only | 9.76 m | 3.03 m | 0.34 m/s | 0.17 m/s |
| Bearing & Doppler | 5.37 m | 2.22 m | 0.21 m/s | 0.13 m/s |

*Table 1. RMSE of position and velocity for fixed and dynamic radars for the scenarios and simulation parameters from Fig.3.*

## Simulation Examples

In this section, we analyze the performance of a DRN in different situations by varying the number of UAVs, their sensing capabilities and the RCS of different targets.

For each Monte Carlo iteration, the UAVs initial positions were randomly generated inside a sphere of radius 30 m at a height of 50 m, the measurement noise was generated according to Gaussian statistics, while the target mobility was modeled according to a random walk. A maximum communication range of 100 m between UAVs and a single hop were considered. For more details about models and parameters, please refer to [15].

In Fig. 3, examples of estimated UAV trajectories for different sensing capabilities have been reported for a specific target path in an environment with NLOS areas. In particular, on the left, examples of DNRs are displayed considering a constant altitude from the ground and the collection of ranging (top) and bearing (bottom) measurements. For comparison, on the right, two situations with a fixed deployment of radar sensors is considered: one with a single terrestrial radar with full sensing capabilities (i.e., capable of retrieving ranging, bearing and Doppler shift information) represented with a diamond in Fig. 3, and another where for fairness the radar network is with the same number (i.e., $N=6$) and sensing capabilities of UAVs.

The ranging error was modeled as $\sigma_r = \sigma_{0r} \cdot d_i^2/\sqrt{\text{RCS}}$, with $\sigma_{0r}$ being the error at the reference distance of 1 m and for a target RCS of 1 m$^2$, and with $d_i$ being the UAV-target distance. In the simulations, the reference ranging error (namely, $\sigma_{0r}$) was set to 0.001 m, while the bearing accuracy was 10 degrees, regardless of the UAV-target distance and RCS. The actual target speed and RCS were set at 1.5 m/s and 0.1 m$^2$, if not otherwise indicated. The UAV trajectories were estimated using a D-optimality criterion and are displayed with dotted grey lines in Fig. 3 and with square markers every 100 time slots. The initial target position is drawn with a black triangle and its actual trajectory with a continuous blue line.

The estimated target trajectory is plotted with a red dashed line with some samples depicted as circles for the same considered time instants. In NLOS propagation conditions, the measurements were not available, and therefore, radars rely on neighbors' collected data or on past estimates.

As we can see, differently from a fixed deployment of radars, having a flying network permits to continuously follow the target with an increased accuracy thanks to the dominant LOS link, even in presence of obstacles. The RMSE results on position and velocity are provided in Table I, showing the superiority of a dynamic configuration. In the case of a single terrestrial radar with full sensing capabilities, the RMSE on position and velocities are of 32.47 m and 0.32 m/s, respectively.

In Fig. 4, the joint impact of the number of UAVs (left) and the target RCS (right) is investigated in terms of averaged RMSE on target position and as a function of the ranging and bearing errors for the dynamic radar configuration. The UAVs were constrained to keep 50 m safety distance from the target. From the results, one can notice that a group of three radars with a millimeter ranging accuracy can obtain the same tracking performance (i.e., approximately 3 meters for target positioning) of three radars with a maximum bearing error of about 13 degrees. Moreover, as expected, we can see that increasing the number of UAVs does not significantly impact the bearing performance, whereas it is beneficial when ranging estimates are used. The effect of the target RCS on the tracking performance is more evident for lower values of $\sigma_{0r}$. The availability of the Doppler information ameliorates the position and velocity estimation accuracy.

## Conclusions and Outlook

In this tutorial, the idea of a UAV DRN for the detection and tracking of malicious UAVs was described. The principles standing behind this concept was presented, with an overview of the system including the description of algorithms that can be adopted by UAVs for target detection and tracking. In contrast with current on-ground radar systems, UAV networks provide new degrees of freedom thanks to their reconfigurability and flexibility. Moreover, the UAVs are considered autonomous in navigating and estimating their best trajectory without impacting the communication latency. The results demonstrate that having a DRN with optimized trajectories, instead of a fixed deployment, helps in preventing NLOS conditions, and, thus, in improved tracking of a passive target.

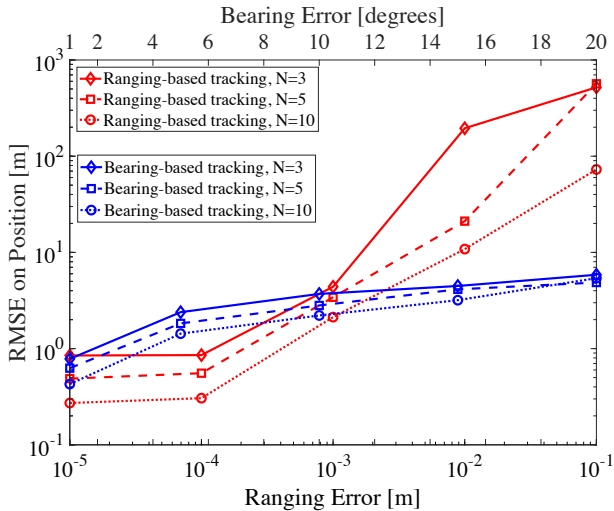 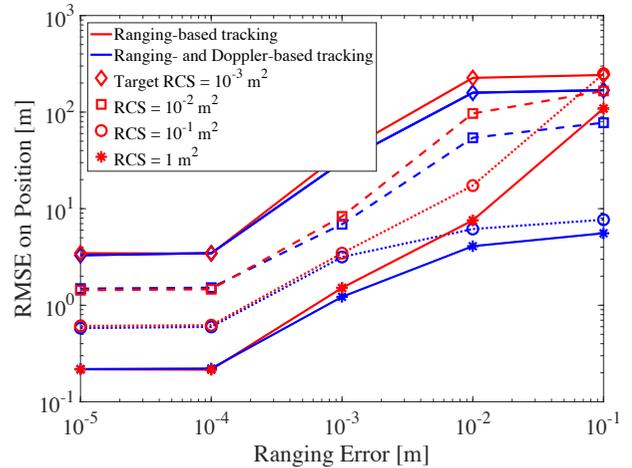

*Figure 4.* Left: RMSE of the target position as a function of the number of UAV-radars. Right: RMSE as a function of the target RCS.

A step forward will be to conceive UAVs with a safety and security payload that can be used as an add-on service in addition to the main primary service for which they are developed (e.g., good delivery or 5G coverage enhancement). This will permit to enhance the safety of cities and the citizen acceptance without the deployment of a dedicated aerial infrastructure. In this way, the considered UAVs will form a multi-functional/multi-sensor network capable both to complete the mission for which each UAV was initially assigned (primary service), and to form a dynamic radar team in order to eradicate possible threats arising from malicious UAVs present in the environment. Finally, machine learning techniques can be designed to improve the navigation performance without burdening the complexity of the processing on-board, for example by exploiting graph neural network solutions.

## References

[1] H. Shakhatreh *et al.*, "Unmanned aerial vehicles (UAVs): A survey on civil applications and key research challenges," *IEEE Access,* vol. 7, pp. 48572--48634, 2019.

[2] [Online]. Available: https://www.sesarju.eu/U-space.

[3] B. Li *et al.*, "Secure UAV Communication Networks over 5G," *IEEE Wireless Commun.,* 2019.

[4] R. Altawy and A. M. Youssef, "Security, privacy, and safety aspects of civilian drones: A survey.," *ACM Trans. Cyber-Physical Sys.,* vol. 1, no. 2, p. 7, 2017.

[5] [Online]. Available: https://www.artsys360.com/.

[6] I. Guvenc *et al.*, "Detection, tracking, and interdiction for amateur drones," *IEEE Commun. Mag.,* vol. 56, no. 4, pp. 75--81, 2018.

[7] V. Clarkson and J. Palmer, "New Frontiers in Passive Radar," *IEEE Potentials,* vol. 38, no. 4, pp. 9--15, 2019.

[8] J. Wang *et al.*, "Taking drones to the next level: Cooperative distributed unmanned-aerial-vehicular networks for small and mini drones," *IEEE Veh. Technol. Mag.,* vol. 12, no. 3, pp. 73--82, 2017.

[9] J. Tisdale, Z. Kim and J. K. Hedrick, "Autonomous UAV path planning and estimation," *IEEE Robotics & Automation Mag.,* vol. 16, no. 2, pp. 35--42, 2009.

[10] P. Hügler *et al.*, "Radar taking off: New capabilities for UAVs," *IEEE Microwave Mag.,* vol. 19, no. 7, pp. 43--53, 2018.

[11] D. Solomitckii *et al.*, "Technologies for efficient amateur drone detection in 5G millimeter-wave cellular infrastructure," *IEEE Commun. Mag.,* vol. 56, no. 1, pp. 43--50, 2018.

[12] D. Dardari, P. Closas and P. M. Djurić, "Indoor tracking: Theory, methods, and technologies," *IEEE Trans. Veh. Technol.,* vol. 64, no. 4, pp. 1263--1278, 2015.

[13] F. Meyer *et al.*, "Distributed localization and tracking of mobile networks including noncooperative objects," *IEEE Trans. Signal Inf. Process. Netw.,* vol. 2, no. 1, pp. 57--71, 2015.

[14] D. Ucinski, *Optimal measurement methods for distributed parameter system identification*, CRC Press, 2004.

[15] A. Guerra *et al.*, "Collaborative target-localization and information-based control in networks of UAVs," in *2018 IEEE 19th Int. Workshop Signal Process. Adv. Wireless Commun. (SPAWC)*, 2018.

**Acknowledgements**

This work has received funding from the European Union's Horizon 2020 research and innovation programme under the Marie Sklodowska-Curie project AirSens (grant no. 793581). P. M. D. thanks the support of the NSF under Award CCF-1618999.

**Anna Guerra** (S'13, M'16) received the M.Sc. degree in electronics and telecommunications engineering and the Ph.D. degree in telecommunications engineering from the University of Bologna, Italy, in 2011 and 2016, respectively. She is a Marie Skłodowska-Curie research fellow within the H2020 European Framework for a project with the University of Bologna, Italy, and the Stony Brook University, New York. Her interests include wireless sensor networks, radio localization, and statistical signal processing. She received the best student paper award at the 2014 IEEE International Conference on Ultra-Wideband in Paris, France.

**Davide Dardari** (M'95 – SM'07) is an Associate Professor at the University of Bologna, Italy. Since 2005, he has been a Research Affiliate at Massachusetts Institute of Technology, USA. His interests are on wireless communications, localization techniques and distributed signal processing.
He received the IEEE Aerospace and Electronic Systems Society's M. Barry Carlton Award (2011) and the IEEE Communications Society Fred W. Ellersick Prize (2012).
He was the Chair for the *Radio Communications Committee* of the IEEE Communication Society and Distinguished Lecturer (2018-2019). He served as an Editor for IEEE TRANSACTIONS ON WIRELESS COMMUNICATIONS from 2006 to 2012.

**Petar M. Djurić** received the B.S. and M.S. degrees in electrical engineering from the University of Belgrade, and the Ph.D. degree in electrical engineering from the University of Rhode Island. He is a SUNY Distinguished Professor and currently a Chair of the Department of Electrical and Computer Engineering, Stony Brook University. His research has been in the area of signal and information processing. Prof. Djuric was a recipient of the EURASIP Technical Achievement Award in 2012. He was the first Editor-in-Chief of the IEEE Transactions on Signal and Information Processing over Networks, and he is a Fellow of IEEE and EURASIP.